\documentclass[preprint]{aastex631}

\submitjournal{AAS Journals}

\usepackage{graphicx}				
\usepackage{latexsym}
\usepackage[T1]{fontenc}
\usepackage{amsmath}
\usepackage{amsfonts}	
\usepackage{amssymb}
\usepackage{scalerel}
\usepackage{lmodern}
\usepackage{gensymb}
\usepackage{mdframed}
\usepackage{silence}

\begin{document}

\title{An upper limit to differential magnification effects in strongly gravitationally lensed galaxies}

\author[0000-0002-0517-7943]{Stephen Serjeant}
\affiliation{School of Physical Sciences \\
The Open University, Milton Keynes, MK7 6AA UK}





\begin{abstract} \label{sec:abstract}
Differential magnification is now well-known to distort the spectral energy distributions of strongly gravitationally lensed galaxies. However, that does not mean that any distortions are possible. Here I prove an analytic upper bound to differential magnification effects. For example, a thermal or sub-thermal CO ladder cannot be made to appear super-thermal just from gravitational lensing, and the Balmer decrement emission line ratio H$\alpha$:H$\beta$ cannot reduce below the case B prediction 
just from differential magnification. In general, if a physical model of a galaxy predicts upper and/or lower bounds to an emission line ratio, then those bounds also apply to the differentially magnified strongly gravitationally lensed case. This applies not just for velocity-integrated emission lines, but also for the line emission in any rest-frame velocity interval. 
\end{abstract}



\section{Introduction}\label{sec:introduction}
The landmark discovery that bright submm and mm-wave galaxies are mainly strongly gravitationally lensed \citep{Negrello+10} has led to a glut of strongly lensed systems \citep[e.g.][]{Negrello+17,Reuter+20,Bears1}. 
Taken in conjunction with optical and near-infrared strong lensing catalogues, there are now of the order a thousand confirmed or candidate strong lensing systems, and future dark energy missions are expected to increase these numbers by two orders of magnitude \citep[e.g.][]{Collett2015}. 
Many spectroscopic or continuum studies currently rely either on low angular resolution data or integrated fluxes across the system \citep[e.g.][]{Bears1,Bears2,Bears3,zGal1,zGal2,zGal3}. Therefore studies have been made to assess the effects of differential gravitational lensing magnification on the observed integrated spectral energy distributions of strongly lensed galaxies 
\citep[e.g.][]{Serjeant2012, Hezaveh+12}. These effects are now widely cited and recognised, but sometimes anomalies in observations are wrongly attributed to differential magnification. This paper therefore sets out to address this by proving an analytic bound on differential magnification effects. 

\section{Method} \label{sec:method}

The formalism is a continuum generalisation of the discrete argument we presented in Appendix E of \cite{Bears3}. 
I consider two emission lines, numbered 1 and 2, and line 2 is always brighter than line 1. For example, line 2 could be the H$\alpha$ emission line, and line 1 could be H$\beta$ under case B recombination with the additional possibility of dust reddening reducing the Balmer decrement. Alternatively, both lines could be $L^\prime$ luminosity measures of CO rotational transitions. 

The line luminosities can be written as $L_1=kL_2$, with 
\begin{equation}\label{eqn:FundamentalBound}
0\leq k\leq 1~.
\end{equation}
This line ratio $k$ can vary across a galaxy, $k=k(x,y)$ where $(x,y)$ are Cartesian coordinates. It immediately follows that a luminosity increment within $\mathrm{d}x\mathrm{d}y$ will satisfy 
\begin{equation}
k(x,y) L_2(x,y) \mathrm{d}x\mathrm{d}y \leq L_2(x,y) \mathrm{d}x\mathrm{d}y ~.
\end{equation}
Integrating over the $(x,y)$ positions in a galaxy image $G$ obtains 
\begin{equation}
\int_G k(x,y) L_2(x,y) \mathrm{d}x\mathrm{d}y \leq \int_G L_2(x,y) \mathrm{d}x\mathrm{d}y ~.
\end{equation}
I define a luminosity-weighted observed mean $k$ as 
\begin{equation}
\langle k \rangle_\mathrm{unlensed}=\frac{\int_G k(x,y) L_2(x,y)\mathrm{d}x\mathrm{d}y}{\int_G L_2(x,y)\mathrm{d}x\mathrm{d}y} \leq 1.
\end{equation}

In other words, integrating the (unlensed) light over a galaxy will not yield a line ratio that exceeds the $k\leq 1$ bound that applies in any individual region. For example, if one measures Balmer line luminosities relative to case B predictions, 
then it follows that it is not possible to make the integrated H$\beta$:H$\alpha$ luminosity ratio exceed the case B recombination limit, if case B recombination applies throughout the galaxy. 

This galaxy is then differentially magnified by a net factor $\mu(x,y)$, where $\mu\geq0$ (i.e. summing the modulus of the magnification factors of each image, so images with negative parity contribute positively to the total magnification of a differential region $\mathrm{d}x\mathrm{d}y$ in the galaxy), and $(x,y)$ now measures positions in the source plane. Again, a differential element in the source plane $\mathrm{d}x\mathrm{d}y$ will satisfy 
\begin{equation}
k(x,y) \mu(x,y) L_2(x,y) \mathrm{d}x\mathrm{d}y \leq \mu(x,y) L_2(x,y) \mathrm{d}x\mathrm{d}y
\end{equation}
(which follows immediately from Eqn.\,\ref{eqn:FundamentalBound}). 
Integrating over the $(x,y)$ positions in a galaxy image $G$ obtains 
\begin{equation}
\int_G k(x,y) \mu(x,y) L_2(x,y) \mathrm{d}x\mathrm{d}y \leq \int_G \mu(x,y) L_2(x,y) \mathrm{d}x\mathrm{d}y ~.
\end{equation}
One can then define a luminosity-weighted observed mean $k$ in the gravitationally-lensed case as 
\begin{equation}
\langle k \rangle_\mathrm{lensed}=\frac{\int_G k\mu L_2\mathrm{d}x\mathrm{d}y}{\int_G \mu L_2\mathrm{d}x\mathrm{d}y} \leq 1~.
\end{equation}

\section{Discussion}\label{sec:discussion}
There is therefore a straightforward hard upper limit to the observed integrated line ratios in a differentially magnified, strongly gravitationally lensed galaxy. 
What can be learned about the underlying line ratios, in the absence of a magnification map $\mu(x,y)$? Firstly, there is no guarantee that any region within the galaxy has a line ratio exactly equalling the observed one. This can be trivially shown to be true even in the absence of differential magnification, i.e. $\mu=$\,constant: if the galaxy is evenly split between regions of $k=0$ and $k=1$, then the average observed $\langle k\rangle=\frac{1}{2}$, even though no $(x,y)$ position will have that line ratio. 

Nevertheless, if the lensed line ratio exceeds $1$, then it must be the case that the fundamental assumption in Eqn.\,\ref{eqn:FundamentalBound} does not hold, i.e. $k(x,y)>1$ in at least part of the galaxy. Therefore, if the integrated H$\beta$ emission line exceeds the case B recombination prediction in a strongly lensed galaxy, then at least part of the galaxy must somehow have a region exceeding that prediction. Similarly, if the lensed CO ladder appears super-thermal, then there must be at least one region within the galaxy that has super-thermal CO transitions. An equivalent argument for line ratio lower bounds can trivially be constructed by exchanging lines 1 and 2 in Section \ref{sec:introduction}. Although this paper describes galaxy lensing, the argument is independent of the structure of the magnification map $\mu(x,y)$, so it applies in all gravitational lensing, whether extragalactic or Galactic. In general, if a physical model predicts lower or upper bounds to a line ratio, then those bounds also apply to the differentially magnified strongly lensed case, so the model can in principle be ruled out even without a magnification map. 

There are examples of extreme, super-thermal CO line transitions in four galaxies from the Bright Extragalactic ALMA Redshift Survey \citep[BEARS,][]{Bears3}. These line ratios can therefore not be attributed solely to differential magnification. These observations are slightly complicated by the fact that the data is from an interferometer, so differences in the Fourier $uv$ coverage between emission lines could in principle lead to an anomalous line ratio, but the $uv$ coverage is thorough in this case so this is not an obvious solution. The argument in this paper holds not just for velocity-integrated emission lines, but also for the line emission in any rest-frame velocity interval; however the emission line profiles are comparable in most of the targets. Deeper follow-up observations are recommended. 


\section{Conclusion}\label{sec: conclusion}
If a physical model of a galaxy predicts upper and/or lower bounds to an emission line ratio, then those bounds also apply to the differentially magnified strongly gravitationally lensed case. This applies not just for velocity-integrated emission lines, but also for the line emission in any rest-frame velocity interval. 

\section{Acknowledgements}
SS acknowledges support from the ELSA project. ``ELSA: Euclid Legacy Science Advanced analysis tools'' has received funding from the European Union’s Horizon Europe research and innovation programme under Grant Agreement no. 101135203. UK participation is funded through the UK Horizon guarantee scheme under Innovate UK grant 10093177. SS thanks Tom Bakx, Steve Eales and George Bendo for constructive conversations on this topic during a workshop with visitors' funding from STFC under grant ST/W000830/1. 


\bibliography{diffmag_rnaas.bib}



\end{document}